\documentclass[fleqn,usenatbib]{mnras}

\usepackage{newtxtext,newtxmath}

\usepackage[T1]{fontenc}
\usepackage{ae,aecompl}

\usepackage{graphicx}	
\usepackage{amsmath}	
\usepackage{amssymb}	
\usepackage{microtype}

\usepackage{xcolor}
\newcommand\correct[1]{\textcolor{black}{#1}}
\newcommand{\df}{NGC 1052-DF2}

\title[On the Rotation of \df]{The Globular Cluster Population of \df: 
Evidence for Rotation}

\author[G. F. Lewis, B. J. Brewer \& Z. Wan]{
Geraint F. Lewis$^{1}$\thanks{E-mail: geraint.lewis@sydney.edu.au (GFL)},
Brendon J. Brewer$^{2}$,
and Zhen Wan$^{1}$
\\
$^{1}$Sydney Institute for Astronomy, School of Physics, A28, The University of Sydney, NSW 2006, Australia\\
$^{2}$Department of Statistics, The University of Auckland, Private Bag 92019, Auckland 1142, New Zealand
}

\date{Accepted XXX. Received YYY; in original form ZZZ}

\pubyear{2019}

\begin{document}
\label{firstpage}
\pagerange{\pageref{firstpage}--\pageref{lastpage}}
\maketitle

\begin{abstract}
Based upon the kinematics of ten globular clusters, it has recently been claimed that the ultra-diffuse galaxy, \df, lacks a significant quantity of dark matter. Dynamical analyses have generally assumed that this galaxy is pressure supported, with the relatively small velocity dispersion of the globular cluster population indicating the deficit of dark matter. However, the presence of a significant rotation of the globular cluster population could substantially modify this conclusion. Here we present the discovery of such a signature of rotation in the kinematics of \df's globular clusters, with a velocity amplitude of $\sim12.44^{+4.40}_{-5.16}$ km/s, which, through Bayesian model comparison, represents a marginally better fit to the available kinematic data;
note that this rotation is distinct from, and approximately perpendicular to, the recently identified rotation of the stellar component of \df.
Assuming this truly represents an underlying rotation, it is shown that the determined mass depends upon the inclination of the rotational component and, with a moderate inclination, the resultant mass to light ratio can exceed $M/L\sim10$. 
\end{abstract}




\section{Introduction} \label{sec:intro}
Recently, \citet{2018Natur.555..629V} reported that the ultra-diffuse galaxy, \df, was lacking in dark matter, with a mass-to-light ratio of order unity, at odds with the dark matter dominance in similar galaxies \citep[e.g.][]{2015ApJ...804L..26V}. This conclusion is drawn from a dynamical analysis of the globular cluster population of \df, showing it to possess an intrinsic velocity dispersion of $\sigma=3.2_{-3.2}^{+5.5}$ km/s, resulting in a dark matter estimate of $\sim3.4\times10^8 \rm{M_\odot}$ within $\sim7.6$ kpc, whilst the stellar mass of \df\ is estimated to be $\sim1-2\times10^8 \rm{M_\odot}$.

This result has proven controversial, with \citet{2018ApJ...859L...5M} questioning the underlying statistical approach,
arguing that the intrinsic velocity dispersion is $\sigma_{\rm int} = 9.5^{+4.8}_{-3.9}$ km/s, implying a mass to light ratio of $M/L_V < 8.1$ at the 90-percent confidence level.
Furthermore, there is an ongoing discussion on the precise distance to \df\ \citep{2018RNAAS...2c.146B,2018ApJ...864L..18V,2019MNRAS.486.1192T,2019RNAAS...3b..29V,2019ApJ...880L..11M}, with suggestions that it might be located at $\sim10$ Mpc,
significantly closer than the $\sim20$ Mpc identified by  \citet{2018Natur.555..629V}, which alleviates the lack of dark matter.
Additionally, others have examined systematic uncertainties due to the adopted approaches to determining the underlying mass distribution, considering simple estimators and more complex modelling \citep{2018MNRAS.481L..59H,2018ApJ...863L..15W,2019MNRAS.484..510N}.
If this galaxy is truly lacking appreciable quantities of dark matter, it existence provides some theoretical challenges within the standard framework for galaxy evolution \citep[e.g.][]{2018ApJ...863L..17N}, as well as for alternative models of gravity \citep[e.g.][]{2018MNRAS.480..473F,2019A&A...626A..47H,2019MNRAS.487.2441H}.

The mass determinations of \df\ have generally relied on the assumption that the globular cluster population is relaxed and pressure supported, and so the velocity dispersion is directly related to the underlying gravitational potential. However, recent observations of the  globular cluster population of our own Milky Way show substantial substructure, retaining the kinematic signature of accretion \citep{2019arXiv190608271M}. This, as well as the discovery of multiple rotational components in the globular cluster population of Andromeda~\citep{2013ApJ...768L..33V,2019MNRAS.484.1756M,2019Nature}, suggests that velocity dispersion mass determinations using globular clusters as simple, pressure-supported tracers might not be appropriate.  
Given this, we revisit the question of whether the globular cluster system of \df\ exhibits a significant rotational signature, potentially demonstrating a more recent accretion origin, and the resultant impact on the inferred dark matter mass.

In this paper,  Section~\ref{model} describes our approach to dynamical modelling and Bayesian model assessment of kinematics with and without a rotational component.
The results of this study are presented in Section~\ref{results}, with the implications for the mass of \df\ discussed in Section~\ref{sec:mass}.
Our conclusions are presented in Section~\ref{conclusions}. In the following analysis, we employ \correct{we employ the original sample presented by \citet{2018Natur.555..629V}, but with the corrected velocity for one of the globular cluster, GC-98, derived from a combination of Keck \correct{Deimos} and LRIS spectra \citep{2018RNAAS...2b..54V}, allowing comparison to previous kinematic studies. 
}

\begin{figure}
    \includegraphics[width=0.9\columnwidth]{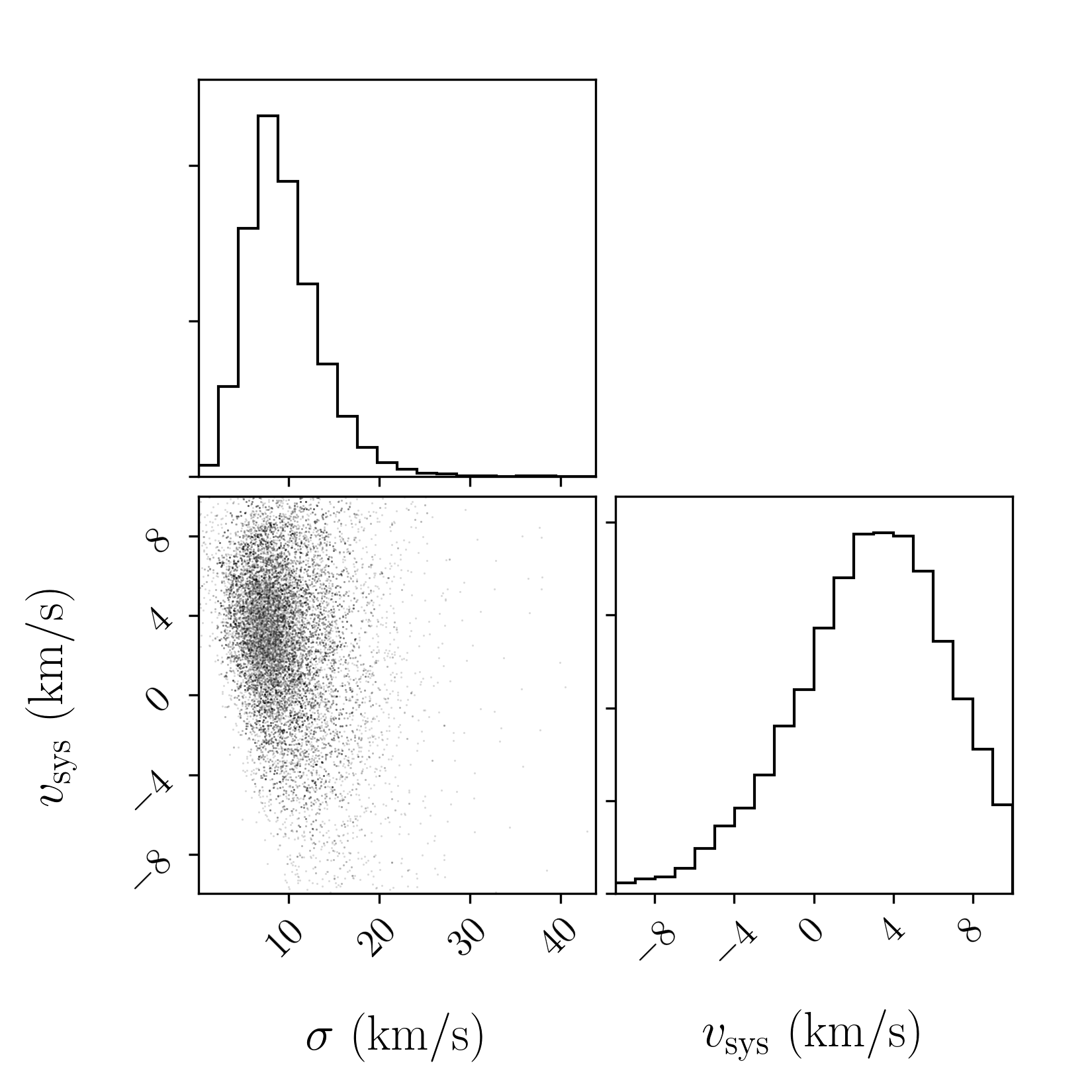}
        \caption{
        The posterior distributions of the systemic velocity, $v_{\rm sys}$, and velocity dispersion, $\sigma$, for the non-rotating model for \df\ presented in Section~\ref{subsec:rot}. These recover the results identified previously \citep{2018Natur.555..629V,2018ApJ...859L...5M}.
        }
        \label{fig:nonrot}
\end{figure}

\section{Dynamical modelling}\label{model}

\subsection{Kinematic Model}\label{subsec:kinematic}

The kinematic model we employ reflects that used in previous studies, comprising of two distinct components, 
one describing the rotation of the system, and the other an underlying dispersion in velocity
\citep[e.g.][]{2001ApJ...559..828C,2013ApJ...768L..33V,2019MNRAS.484.1756M,2019Nature}. The rotational velocity is of the form
\begin{equation}
    v_{\rm rot}(\theta) = A \sin\left( \theta - \phi \right)
    \label{rotation}
\end{equation}
where $A$ is the amplitude of the rotation, and $\phi$ represents the axis of rotation in the chosen coordinate system\correct{; note that, for simplicity, $A$ is assumed to be constant with no radial dependence.}
The velocity dispersion is assumed to be constant and is represented by $\sigma$. It is assumed that the rotation is centred on a systemic velocity which is also treated as a free parameter and is denoted $v_{\rm sys}$; note that for the analysis presented in this paper, a constant value of $1800$ km/s was subtracted from all velocities, and our $v_{\rm sys}$ is relative to this value.
Clearly, if $A=0$ in Equation~\ref{rotation}, the velocity distribution of globular clusters will be described entirely by the velocity dispersion, $\sigma$, and so will correspond to previous models employed in describing \df\ \citep[e.g.][]{2018Natur.555..629V,2018ApJ...859L...5M}. Hence, the models considered in this paper are generalizations of previous models
with the dispersion-only case being a special case of our model.

\subsection{Bayesian Inference}\label{subsec:bayesian}
Throughout this letter we use the Bayesian framework  to describe uncertainties
using probabilities and probability distributions \citep{o2004kendall,gregory2005bayesian}, noting that this approach has been used \correct{previously} in assessing the veracity of dynamical \correct{models of stellar} systems \citep{2014MNRAS.437.3172D,2019Nature}. 

The posterior distribution for
parameters $\Lambda$ given data $D$ is given by
\begin{align}
    p(\Lambda | D) &= \frac{p(\Lambda)p(D | \Lambda)}{p(D)}
\end{align}
where $p(\Lambda)$ is the prior distribution, $p(D | \Lambda)$ the likelihood function,
and $p(D) = \int p(\Lambda)p(D|\Lambda)\,d\Lambda$ is the marginal likelihood (also known
as the `evidence' or `Bayesian evidence').
The scientific question of interest primarily
involves the value of $A$, the amplitude of any rotation in \df.

For the sample of globular clusters in \df\ described by a position $(r_i,\theta_i)$ and velocity with associated error $(v_i,s_i)$,
the kinematic model described in Section~\ref{subsec:kinematic}, we can define the likelihood function by
\begin{equation}
    {\cal L} = \prod_i 
    \frac{1}{\sqrt{ 2 \pi \sigma_i^2 }} 
    \exp
    \left( - \frac{(v_i - ( v_{\rm rot}(\theta_i) + v_{\rm sys} ))^2}{2 \sigma_i^2} \right)
    \label{likelihood}
\end{equation}
where the product is over the individual data points and 
\begin{equation}
    \sigma_i^2 = \sigma^2 + s_i^2.
    \label{sigma}
\end{equation}
Note that for this study, we do not consider the question of potential contamination of the globular cluster population by interlopers \citep[e.g. see][]{2018ApJ...859L...5M}.

We use DNest4 \citep{Brewer2011, JSSv086i07} to explore the posterior distribution and calculate the marginal likelihood for each of the models considered. Table~\ref{tab:prior} outlines the prior distributions employed for the parameters considered in this study. These are simple
prior uniform priors with moderate bounds on the parameters. To allay any concerns about
the sensitivity of Bayesian model selection to the priors, in Section~\ref{subsec:bjb_priors}
we present an analysis with non-uniform informative priors and no model comparison.

\begin{table}
    \centering
    \begin{tabular}{|l|ll|}
    \hline
    Parameter & Prior & Unit \\
    \hline
      $A$        &   Uniform$(0,20)$   & km/s  \\
      $\phi$ &   Uniform$(0,2\pi)$ & radians \\
      $\sigma$  &   Uniform$(0,20)$   & km/s  \\
      $v_{\rm sys}$  &   Uniform$(-10,10)$ & km/s \\
      \hline
    \end{tabular}
    \caption{The prior probability distributions employed in this study. Note that the systemic velocity is relative to $1800$ km/s.
    }
    \label{tab:prior}
\end{table}

\section{Results}\label{results}

\subsection{Non-Rotation}\label{subsec:nonrot}
As noted previously, neglecting the rotational component, so the kinematics are represented by only the velocity dispersion, should recover previous analyses of \df\ \citep[e.g.][]{2018ApJ...859L...5M}. In Figure~\ref{fig:nonrot} we present the resultant posterior
distributions of $\sigma$ and $v_{\rm sys}$ as a corner plot \citep{corner}, and it is clear that the previous results have been recovered, with a velocity dispersion of
$\sigma = 8.96^{+4.32}_{-3.08}$ km/s
and a systematic velocity of
$v_{\rm sys} = 3.05^{+3.47}_{-4.06}$ km/s.
The estimate of the marginal likelihood for this model is $\log(Z) = -40.21$.   
\subsection{Rotation}\label{subsec:rot}

\begin{figure}
    \includegraphics[width=\columnwidth]{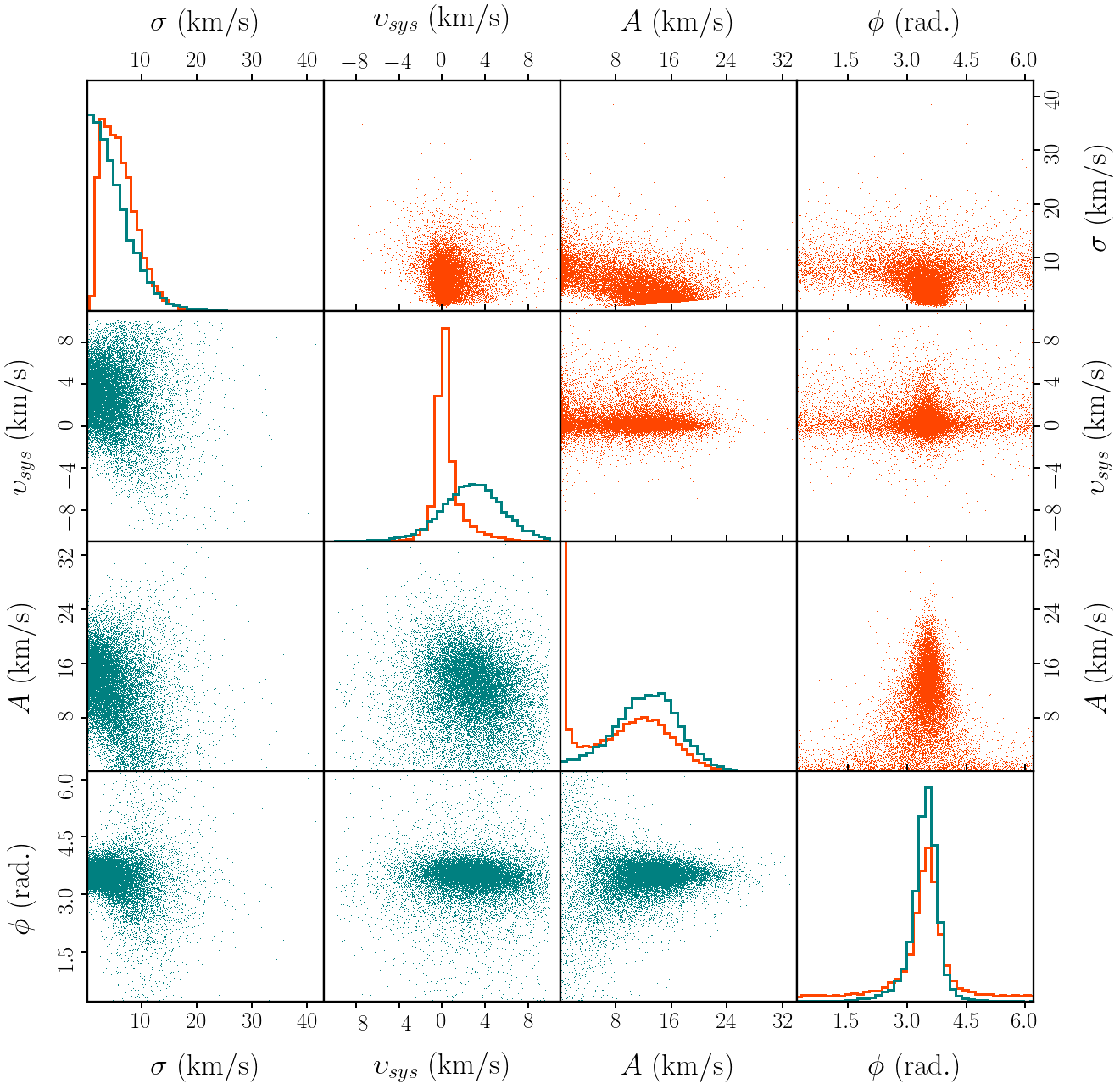}
        \caption{As Figure~\ref{fig:nonrot}, but now for the parameters in the case the rotational component is considered. The blue points and histograms (lower-left) corresponds to the uniform priors discussed in Section~\ref{subsec:rot}, whereas the red points and histograms (upper-right) corresponds to the alternative priors presented in Section~\ref{subsec:bjb_priors}. In the latter case, the posterior for $A$ is bimodal with
        some probability of a tiny value and some probability of a more substantial value, and the joint posteriors have also been affected by the
        non-uniform priors in the expected ways.
        }
        \label{fig:rot}
\end{figure}

\begin{figure}
    \includegraphics[width=0.9\columnwidth]{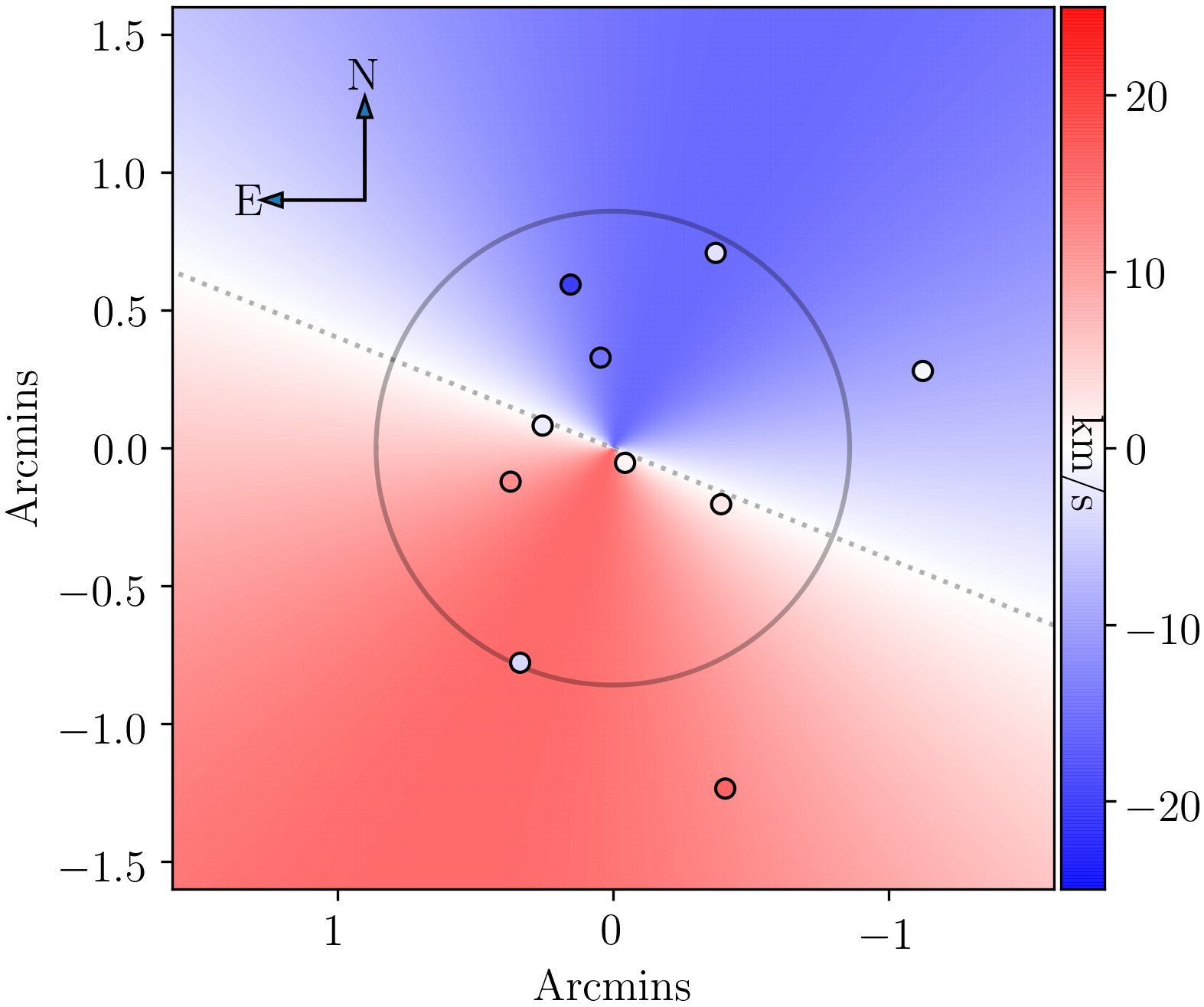}
        \caption{An illustration of the rotational component of the best fit model for \df\ as described in Section~\ref{subsec:rot}. \correct{Note that the model angle, $\theta$, is defined as from East towards North, whereas the standard astronomical Position Angle, PA, is from North towards East.}
        }
        \label{Fig:df2}
\end{figure}

The question of the presence of a rotational component of the globular cluster population was considered by \citet{2018Natur.555..629V} who concluded, after examining velocities along the major and minor axes of \df, that no rotational signature is present. 
The corner plot summarising the results of this exploration of the posterior distribution of the modelling including the rotational component is presented 
as the blue points and histograms presented in the lower-left of
Figure~\ref{fig:rot}. The model
favours an amplitude $A$ that is significantly above zero, with a suppression of the velocity dispersion, $\sigma$.
The parameters are inferred to be
$A=12.44^{+4.40}_{-5.16}$ km/s, $\phi=3.47^{+0.28}_{-0.35}$ radians,
$\sigma=3.88^{+4.42}_{-2.72}$ km/s, and $v_{\rm sys}=2.90^{+2.67}_{-2.83}$ km/s; note that the angle here is taken as from East through North.
These values are the posterior median and a 68\% central credible interval, but it should be noted that 
$\sigma$ is highly asymmetric and is heavily skewed towards zero.
An upper ``limit'' (really the $68^{th}$ and $95^{th}$ percentiles of the posterior distribution) for $\sigma$
is 5.70 km/s and 12.11 km/s respectively.
For illustration purposes, we have also plotted the rotational model for the
maximum likelihood parameters in Figure~\ref{Fig:df2}. 

\citet{2018Natur.555..629V} examined the structure parameter for the underlying stellar distribution of \df\ finding it to be slightly elliptical with a minor axis appears to be at approximately $\theta\sim3.9$ radians in the above coordinate system. This orientation is similar to the axis of rotation for the globular cluster population identified in the above modelling. 
We note that recent spatially resolved spectroscopy of \df\ with the MUSE spectrograph on the VLT identified a velocity gradient of the stellar component consistent with the presence of significant rotation~\citep{2019A&A...625A..76E,2019A&A...625A..77F}. 
Intriguingly, the maximal velocity gradient is identified at a $PA\sim30^o$, with the conclusion that the stellar body of \df\ represents a prolate rotator. This implies that the axis of rotation of the globular cluster population of \df\ identified in this paper, which is approximately along the minor axis axis, is consistent with being perpendicular to that of the stellar component.   
However, it is worth noting that if the motions of globular cluster population of \df\ represent the dynamical memory of an accretion event, there there is no reason {\it a priori} that we should expect its rotational axis to be aligned with the structural form of the stellar component, a situation observed in the recent study of Andromeda, where neither rotation signature identified in the M31 globular clusters is aligned with the underlying properties of the spiral galaxy~\citep{2019Nature}. 
From this, we conclude that the kinematics of the globular cluster population of \df\ is distinct from stellar component, potentially indicating a distinct origin.
\correct{We also note that the presence of this rotation might indicate a relatively recent accretion of the globular cluster system of \df, suggesting that the population is not necessary relaxed, hence might provide a poor determinant of the underlying mass. This potentially explains the discrepancy with mass determination from the stellar population \citep[e.g.][]{2019A&A...625A..76E}.}

The resultant Bayesian evidence for this exploration is $\log(Z) = -39.77$.
This means that this rotational model, even though it has two additional parameters compared to the pure velocity dispersion model considered in Section~\ref{subsec:rot}, presents a better description of the data than the non-rotating model~\footnote{\correct{
We note that two additional globular cluster candidates, GC-93 and T3, have been identified by \citet{2019A&A...625A..76E} and \citet{2019MNRAS.486.1192T} resp. The analysis was repeated including these, finding $\log(Z)=-48.76$ for the non-rotating model, and $\log(Z)=-48.19$ for the rotating model. Hence the addition of these two additional clusters do not significantly change the ratio of the evidences, with the rotating model still favoured over the non-rotating model.
}}.
However, the difference is not overwhelming. If the prior probabilities
for no-rotation and rotation are $1/2$ each, the posterior probabilities
are 0.39 and 0.61 respectively. The marginal likelihood for the disjunction of
these two models (i.e., the probability density of the data given that either
non-rotation or rotation is true) is $\log(Z)=-39.97$.

\subsection{Alternative priors}\label{subsec:bjb_priors}
The above analysis involved a model comparison calculation where the parameters had uniform
priors. The bounds of the priors were sensible, but we should take care in interpreting the
results because the marginal likelihoods would change if the prior bounds were changed.
In this subsection, we perform an analysis with the rotation model using informative but
heavy-tailed priors that allow the amplitude $A$ to be potentially very small but not
precisely zero. The joint prior for the parameters is:
\begin{align}
    \log_{10}\sigma             &\sim \textnormal{Student-}t(1, 0.5, 4) \\
    v_{\rm sys} \,|\, \sigma    &\sim \textnormal{Student-}t(0, 0.1\sigma, 1) \\
    t                           &\sim \textnormal{Student-}t(0, 2, 2) \\
    A                           &\leftarrow \left(10^{1 - |t|}\right)\sigma
\end{align}
where the three parameters of the Student-$t$ distribution are the location, scale,
and shape (degrees of freedom) respectively. The intuitions behind these choices are
as follows.

Firstly, the prior median for $\sigma$ is 10 km/s, similar to the scale identified previously, but the overall prior is broad with
the $(2.5, 16, 50, 84, 97.5)$ percentiles being $(0.41, 2.71, 10.00, 36.92, 244.59)$ km/s respectively.
A log-normal distribution could capture this but the Student-$t$ with
shape parameter 4 has heavier tails and is therefore more fail-safe. If $\sigma$ were
known, this would influence the plausible orders-of-magnitude for $v_{\rm sys}$ and
$A$, so their priors are dependent on $\sigma$. Firstly, $v_{\rm sys}$ is probably
small in absolute value compared to $\sigma$ and can be positive or negative (though
there is some chance it is of comparable magnitude to $\sigma$, with the heavy tails), and
$A$ might be comparable to $\sigma$, a little less, or a lot less (potentially several
orders of magnitude).
The corner plot based on these priors is presented
as the red points and histograms in the upper-right of Figure~\ref{fig:rot}.
The posterior distribution for $A$ has a peak near zero which contains about
30\% of the posterior probability, \correct{reflecting the non-rotating model},
and a broad peak around 10 km/s which contains
the remaining 70\%, \correct{representing the case with rotation}. The posterior probability that $A$ > 5 km/s is 64\% and the
prior probability of this statement was 39\%. This supports the simple-prior analysis
\correct{which was presented in Section~\ref{subsec:nonrot} and \ref{subsec:rot}}
in that it suggests there is evidence for a rotation in \df,
but it is not overly conclusive.

The inferred value of the orientation $\phi$ of the potential rotation is
consistent with the simple prior analysis, and the other parameters' posterior
distributions have been affected by the non-uniform priors to put more probability
near zero.
The marginal likelihood with these alternative priors is $\log(Z) = -38.81$.

\section{Implications for Mass}\label{sec:mass}

As noted in the introduction, several  estimators have been employed in determining the mass of \df, typically assuming that the system is pressure supported, although we note that \citet{2019MNRAS.484..510N} did consider the question of a rotational component in their distribution function modelling.
If the signature identified in Section~\ref{subsec:rot} does indicate the presence of a significant rotation component, this will influence the derived dynamical mass of \df. 
To illustrate this, we consider a simple toy estimator for the total mass given by
\begin{equation}
    M(<r) = \left( \left( \frac{v_{\rm rot}}{\sin(i)} \right)^2 + \sigma^2 \right) \frac{r}{G}
    \label{mass}
\end{equation}
where $v_{\rm rot}$ represents the rotational velocity, $\sigma$ the velocity dispersion, and 
$i$ is the inclination of the rotation in the standard astronomical definition. 
Note that more complex mass estimators typically possess a multiplicative constant combined with the square of the velocity dispersion which accounts for the interior mass distribution and
typically exceeds unity, so Equation~\ref{mass} can be taken as a lower estimate of the total mass \citep{2009ApJ...704.1274W,2010MNRAS.406.1220W}. 

\begin{figure}
    \includegraphics[width=0.95\columnwidth]{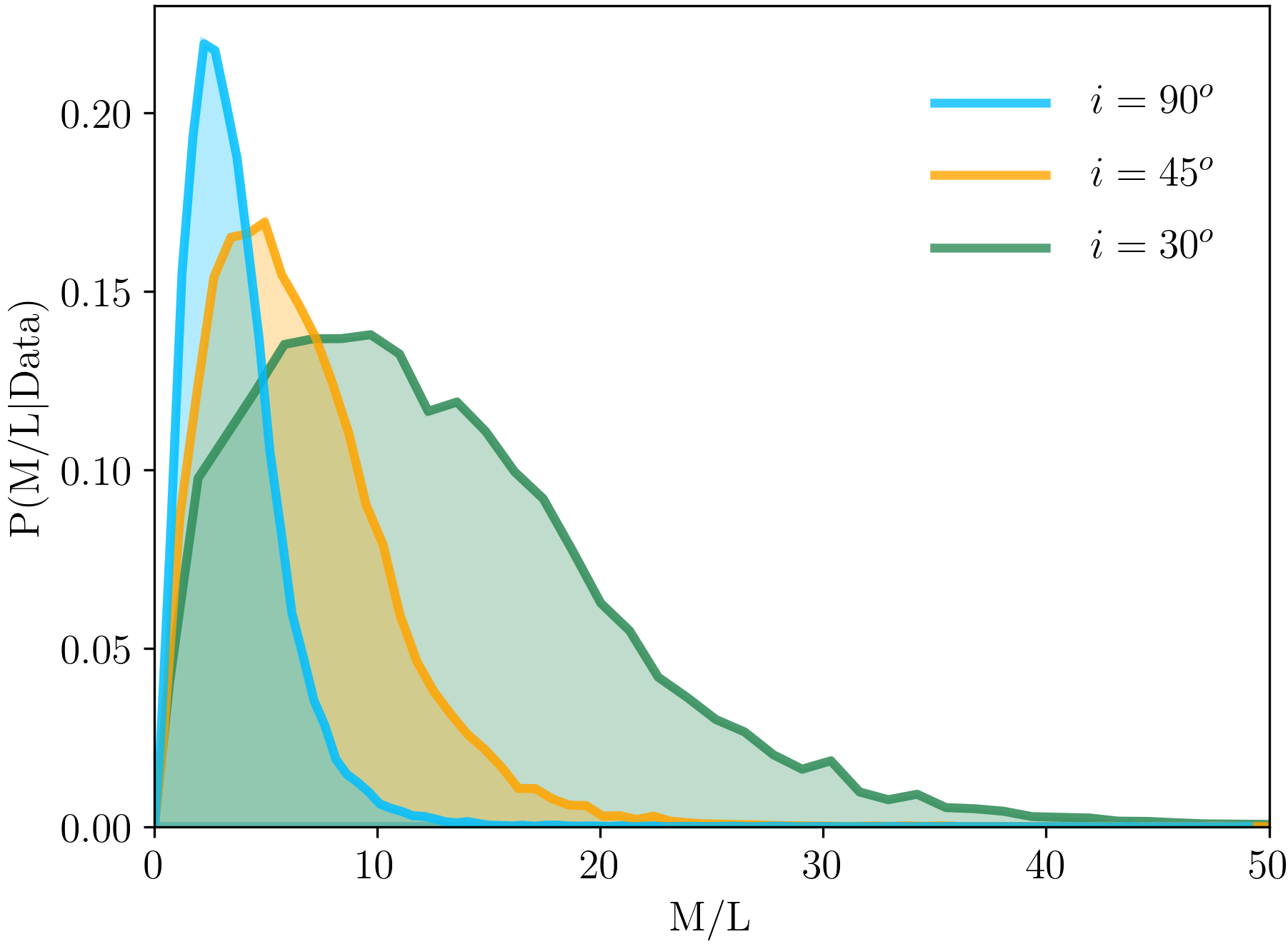}
        \caption{The mass to light ratio of \df\ given the kinematic modelling presented in Section~\ref{sec:mass}. Here, the blue curve corresponds to an inclination of $90^o$, orange an inclination of $45^o$ and an inclination of $30^o$ represented in green.}
        \label{fig:df2-ml}
\end{figure}

Taking the magnitude of the rotational velocity, $A$, in Equation~\ref{rotation} to represent $v_{\rm rot}$, and $\sigma$ to represent the velocity dispersion, we can use the chains of the Monte Carlo exploration to determine the distribution of total mass. In Figure~\ref{fig:df2-ml}, we present the calculated mass to light ratio, $M/L$, assuming a fiducial radius of $r=7.5$ kpc and adopting a stellar mass of $\sim10^8 \rm{M_\odot}$ \citep{2018ApJ...859L...5M}, with the blue representing an inclination of $90^o$, orange an inclination of $45^o$ and an inclination of $30^o$ in green\correct{; note that if the plane of rotation is randomly orientated, there is a 70\% probability of an inclination of $>45^o$ and an 87\% probability of an inclination of $>30^o$}. As seen, as the inclination decreases, and remembering that the mass estimator in Equation~\ref{mass} corresponds to a lower bound, the resultant mass to light ratio becomes substantial, with exceeding $M/L \sim 10$ being quite probable; at an inclination of $i=90^o$, the probability of $M/L>10$ is only $\sim2\%$, but this increases to $\sim20\%$ at $i=45^o$ and $\sim55\%$ at $i=30^o$.
Clearly, the presence of a rotational component in the globular cluster population reduces the requirement that \df\ lacks a substantial quantity of dark matter. 

\section{Conclusions}\label{conclusions}
In this paper, we've re-examined the kinematics of the globular cluster population of \df, considering the possibility of the presence of a significant rotational component. As outlined in Section~\ref{results}, based upon Bayesian model comparison, the existence of a rotational component is slightly more plausible than a model possessing only a velocity dispersion. The presence of this rotation, discussed in Section~\ref{sec:mass}, can significantly influence the inferred mass to light ratio, with a simple mass estimator suggesting that a $M/L > 10$ is probably likely for moderate inclinations, bringing its dark matter content inline with similar galaxies. \correct{
However, even with this larger mass, the relative richness of the globular cluster population ensure \df\ remains peculiar by more than an order of magnitude \citep[see][]{2018MNRAS.481.5592F}.}

Of course, one argument against the analysis presented in this paper is that we are dealing with small number statistics and that we are pushing the data too hard, or even torturing it, by employing a model with additional parameters to extract the dynamical properties and underlying mass distribution \citep[e.g.][]{2019MNRAS.484..245L}. However, as noted, the application of Bayesian evidence naturally encompasses Occam's razor \citep[][Chapter 28]{mackay2003information}, and hence the model comparison presented in this paper robustly demonstrates that the rotational model represents a plausible description of the underlying data, and cannot be simply ruled out.
But clearly more data and more in-depth dynamical modelling is required to fully address the question of where \df\ truly lacks dark matter.

\section*{Acknowledgements}
We thank Foivos Diakogiannis for insightful discussions on mass distributions in stellar systems, and Dougal Mackey and Annette Ferguson for details of globular cluster populations.
GFL acknowledges support from a Partnership Collaboration Award between the University of Sydney and the University of Edinburgh, and thanks the Royal Observatory, Edinburgh for their hospitality where the idea for this study was born.
ZW is supported by a Dean's International Postgraduate Research Scholarship at the University of Sydney.




\bibliographystyle{mnras}
\bibliography{paper} 


\bsp	
\label{lastpage}
\end{document}